\documentclass[a4paper,11pt]{article}

\usepackage{pos}
\usepackage{physics}
\usepackage{graphicx}
\usepackage{subfig}

\newcommand{\at}[2][]{#1|_{#2}}

\title{QCD topology with electromagnetic fields and the axion-photon coupling}
\ShortTitle{QCD topology with EM fields and the axion-photon coupling}

\author[a]{Bastian Brandt}
\author[b]{Francesca Cuteri}
\author[a]{Gergely Endr\H{o}di}
\author*[a]{Jos\'e Javier Hern\'andez Hern\'andez}
\author[a]{Gergely Mark\'o}

\affiliation[a]{Universit{\"a}t Bielefeld,\\
  Universit{\"a}tsstra{\ss}e 25, 33615 Bielefeld, Germany}

\affiliation[b]{Institute for Theoretical Physics, Goethe-Universit{\"a}t\\
Max-von-Laue-Stra{\ss}e 1, 60438 Frankfurt am Main, Germany}

\emailAdd{brand@physik.uni-bielefeld.de}
\emailAdd{cuteri@itp.uni-frankfurt.de}
\emailAdd{endrodi@physik.uni-bielefeld.de}
\emailAdd{hernandez@physik.uni-bielefeld.de}
\emailAdd{gmarko@physik.uni-bielefeld.de}

\abstract{The introduction of non-orthogonal electric and magnetic fields in the QCD vacuum enhances the weight of topological sectors with a nonzero topological charge. For weak fields, there is a linear response for the expectation value of the topological charge. We study this linear response and relate it to the QCD correction to the axion-photon coupling. We also analyse the magnetic field dependence of the topological susceptibility for a range of temperatures around $T_c$. In this work we use lattice simulations with improved staggered quarks at physical masses, including background magnetic and (imaginary) electric fields.}

\FullConference{%
  The 39$^{th}$ International Symposium on Lattice Field Theory (Lattice 2022) \\
  8-13 August 2022 \\
  Bonn, Germany
}

\begin{document}

\maketitle

\section{Introduction}\label{introduction}
The Standard Model of particle physics (SM) is the most accurate description of nature at the microscopic level that humankind has been able to develop so far. These past 50 years have added robustness to the theoretical structure of the SM through several experimental discoveries, such as the confirmation on the existence of quarks, the prediction of certain properties of the weak neutral currents or the recent discovery of the Higgs boson. Despite these and many other remarkable findings, there are still several gears in the SM machinery that we do not yet understand and perhaps are even missing. One particular type of problems stems from a lack of `naturalness', such as the strong CP problem \cite{hook2018tasi}. This last problem is deeply related to axions and the topology of Quantum Chromodynamics (QCD) that we will be discussing in this proceeding.

QCD, a subset of the SM, is a non-Abelian gauge theory with CP symmetry. Nonetheless, Lorentz and gauge invariance as well as renormalisability would allow the addition of a CP-odd term to the QCD Lagrangian, the so-called $\theta$-term, which exclusively depends on the gauge fields. If one indeed includes this term, a nonzero value for the electric dipole moment of the neutron $d_n$ appears, which is severely constrained experimentally, $\abs{d_n} < 1.8 \times 10^{-26}\,e\cdot$cm~\cite{Abel:2020pzs}. Hence, this CP-odd term is suppressed by the $\theta$ parameter, which several lattice studies have bounded as $\abs{\theta} < 10^{-10}$ \cite{Bhattacharya:2021lol,Alexandrou:2020mds,Dragos:2019oxn}. The fact that this parameter is unnaturally small is the strong CP problem.

Many solutions to the strong CP problem in QCD have been suggested, such as the existence of at least one massless quark \cite{PhysRevLett.37.8}
 or the suppression of the $\theta$ parameter due to screening effects from confinement \cite{Ellis:1978hq}. Experimental refutation or the appearance of undesired physical properties rule out many of those suggestions. From the subset of physically allowed solutions, the one proposed by Peccei and Quinn \cite{PhysRevLett.38.1440, PhysRevD.16.1791} is probably the simplest, and thus it has attracted most attention, both theoretically and experimentally in recent times.

The Peccei-Quinn solution postulates the existence of a new U(1)$_{\text{PQ}}$ (anomalous) symmetry that is spontaneously broken in QCD. Associated with this spontaneous symmetry breaking, there is a pseudo-Nambu-Goldstone boson which has been labelled \textit{axion}. The axion only couples directly to a CP-odd combination of the gauge fields, whereas its couplings to the rest of the particles of the SM depend on the specific model (i.e.\ KSVZ \cite{PhysRevLett.43.103, Shifman:1979if} or DFSZ \cite{Dine:1981rt}) and appear only through derivative terms. For the possible detection of axions in contemporary experiments, its most important coupling is the one to photons. The QCD corrections to this coupling are independent of the specific model and are substantial, as was shown using chiral perturbation theory (ChPT)~\cite{di2016qcd}. 

In this proceeding we will show that it is possible to calculate the QCD correction to the axion-photon coupling with lattice QCD by means of simulations at nonzero background magnetic and electric fields. We will present preliminary results for this coupling as well as for the dependence of the topological susceptibility with the magnetic field at finite temperature.

\section{QCD topology with electromagnetic fields}\label{topology}
The vacuum of QCD (in Euclidean space) can be classified into distinct topological sectors characterized by the integer-valued topological charge $Q_{\rm top}$ of the $\mathrm{SU}(3)$ gauge fields,
\begin{equation}
    Q_{\rm top}=\int d^4x\,q_{\text{top}}, \qquad q_{\rm top}=\frac{g^2}{32\pi^2}
    G^b_{\mu\nu}\Tilde{G}_b^{\mu\nu}\,, 
\end{equation}
where $q_{\text{top}}$ is the topological charge density, $G^b_{\mu\nu} = \partial_{\mu}A^b_{\nu} - \partial_{\nu}A^b_{\mu} -f^{bcd}A^c_{\mu}A^d_{\nu}$ is the gluonic field strength and $\Tilde{G}^b_{\mu\nu} = \frac{1}{2}\epsilon_{\mu\nu\rho\sigma} G^{b,\,\rho\sigma}$ its dual\footnote{We use the convention $\epsilon_{0123}=1$.}. The vacuum sectors with different topological charges are distinct in the sense that a smooth continuous transformation cannot change the value of $Q_{\text{top}}$ of the field configuration. 

The coupling of the axion $a$ to gluons is of the form $a/f_a\cdot Q_{\rm top}$ in the action, where $f_a$ is the associated energy scale, the axion decay constant. Thus, for a homogeneous axion field the prefactor can be interpreted as the $\theta$-parameter; $\theta=a/f_a$. 
Moreover, the second moment of $Q_{\text{top}}$, the topological susceptibility $\chi_{\text{top}}$, is related to the mass $m_a$ of the axion as
\begin{equation}
    \chi_{\text{top}} = \frac{\partial^2}{\partial^2 \theta}\ln Z(\theta)\at[\bigg]{\theta = 0} = f_a^2 \frac{\delta^2}{\delta^2 a}\ln Z(a) \at[\bigg]{a = 0} = m_a^2 f_a^2,
\end{equation}
where $Z$ is the partition function of QCD. Hence, an analysis of the temperature dependence of the topological susceptibility can provide information about the cosmological history of the axion.
The topological susceptibility can be calculated both in chiral perturbation theory (see, e.g., Ref.~\cite{di2016qcd}) and on the lattice (see, e.g., Refs.~\cite{Bonati2016axion,Borsanyi2016lattice,Jahn:2018dke}). Together with temperature-effects, the impact of background magnetic fields on $\chi_{\rm top}$ are expected to be relevant for off-central heavy-ion collision phenomenology, in particular in connection with the chiral magnetic effect~\cite{Fukushima:2008xe}, which arises due to a combination of magnetic fields and topology. So far, the $B$-dependence of $\chi_{\rm top}$ has only been calculated in chiral perturbation theory~\cite{Adhikari:2021lbl}.

Besides gluons, the axion also couples to photons. This interaction is analogous to the gluonic one above, this time involving the topological charge $Q_{\rm EM}$ of the photon field. Thus, 
if we introduce background electromagnetic fields $F_{\mu\nu}$ in QCD with nonzero topology -- i.e., with $\Vec{E}\cdot \Vec{B} \neq 0$ -- this becomes equivalent to having an effective $\theta$-term in the action~\cite{DElia:2012ifm}. This generates a nonzero value for the expectation value of $Q_{\text{top}}$ that can be measured on the lattice. As a consequence of $Q_{\text{top}}$ being a CP-odd operator, it can only be an odd function of $\Vec{E}\cdot \Vec{B}$. Hence, for small fields we expect a linear relation between the topological charge and $\Vec{E}\cdot \Vec{B}$. 

We can understand this behaviour further through the Atiyah-Singer index theorem \cite{indexth}. The electromagnetic topological charge induced by a nonzero $\Vec{E}\cdot \Vec{B}$ contributes to the zero modes of the Dirac operator as
\begin{equation}
    n_+ - n_- = Q_{\text{top}} + Q_{\text{EM}} =  \frac{g^2}{32\pi^2} \int d^4x \, G^b_{\mu\nu}\Tilde{G}_b^{\mu\nu} + \frac{e^2}{16\pi^2} \int d^4x \, F_{\mu\nu}\Tilde{F}^{\mu\nu}\,,
\end{equation}
where $e$ is the elementary electric charge and the dual field strength $\Tilde{F}_{\mu\nu}$ is defined analogously to the gluonic case above. Since the zero modes are heavily suppressed in the QCD path integral by the fermion determinant, gluon field configurations whose topological charge cancel the electromagnetic one will be favoured. Hence, we expect opposite signs for the QCD and electromagnetic topological charges. Also, this implies that for weak fields the coefficient of proportionality between $\langle Q_{\text{top}} \rangle$ and $\Vec{E}\cdot \Vec{B}$ must be negative. In the next section we will argue that this coefficient is related to the axion-photon coupling.

\section{The axion-photon coupling} \label{coupling}%

The axion-photon coupling is of the form $g_{a\gamma\gamma} = g_{a\gamma\gamma}^0 + g_{a\gamma\gamma}^{\text{QCD}}$, where the first term is a coefficient depending on the details of the axion model, while the second one depends exclusively on QCD and is model-independent.
The model dependent coefficient is proportional to the ratio of the electromagnetic and colour anomalies. In ChPT, the next-to-leading order (NLO) prediction for the QCD correction is $g_{a\gamma\gamma}^{\text{QCD}}f_a/e^2 = -0.0243(5)$ \cite{di2016qcd}. From now on we will only discuss the QCD correction to the coupling, hence we will omit the QCD superscript. 

The direct coupling between axions and photons is of the form $g_{a\gamma\gamma}^0 aF_{\mu\nu}\Tilde{F}^{\mu\nu}/4=g_{a\gamma\gamma}^0 a\Vec{E}\cdot \Vec{B}$. After the QCD path integral, the QCD corrections to this term in the effective action can therefore be found as 
\begin{equation}
    g_{a\gamma\gamma} f_a = \frac{T}{V}\frac{\partial^2}{\partial \theta\, \partial (\Vec{E}\cdot \Vec{B})} \ln{Z} \, \at[\bigg]{\theta=\Vec{E}=\Vec{B}=0},
\end{equation}
where $Z$ includes constant background electromagnetic fields and 
we again traded a functional derivative with respect to the homogeneous axion field to a derivative with respect to $\theta$.
This relation suggests several methods to extract the coupling, by computing one or two derivatives of the partition function. We will refer to these methods as the electric and the correlator method, respectively. 

For the electric method we only perform the derivative with respect to the $\theta$ parameter. As discussed in Sec.~\ref{topology}, for constant and weak electromagnetic fields, we obtain
\begin{equation}
    \frac{T}{V}\langle Q_{\text{top}} \rangle_{E,B} = g_{a\gamma\gamma}f_a \Vec{E}\cdot \Vec{B}.
\end{equation}

For the correlator method, we also take the derivative with respect to the electric field. Without loss of generality, we take both $\Vec{E}$ and $\Vec{B}$ to point in the $x_3$ direction and under the same assumptions we find
\begin{equation}
    -i\int d^4x\,\langle j_4(x) q_{\rm top}(0) \rangle \, x_3 = g_{a\gamma\gamma}f_a B,
    \label{eq:corrmeth}
\end{equation}
where $j_4$ is the time component of the Euclidean electric current and $q_{\rm top}$ is the topological charge density defined in Sec.~\ref{topology}. 
To arrive at this result, we replaced the homogeneous electric field by an oscillatory one, $E \sin(p_3x_3)$, realised by a photon vector potential $A_4=E/p_3 \cos(p_3 x_3)$. This induces a similarly oscillating topological charge density, which, after differentiating with respect to the amplitude of the electric field gives rise to a momentum-space topology-current correlator. Its zero momentum limit gives Eq.~\eqref{eq:corrmeth}. Oscillatory electric fields of this type were investigated in Refs.~\cite{Endrodi:2021qxz,Endrodi:2022wym} and similar correlators were also used to discuss local CP violation in relation with the chiral magnetic effect~\cite{Bali:2014vja} and to compute the quark spin polarisation at nonzero magnetic fields~\cite{Bali:2020bcn}. 

In summary, by analysing the functional behaviour of $\langle Q_{\text{top}}\rangle$ as a function of $\Vec{E}\cdot \Vec{B}$, and of $\langle j_4 q_{\text{top}}\rangle$ as a function of $\Vec{B}$, we can obtain $g_{a\gamma\gamma}f_a$ via numerical differentiation. From now on we will measure the electromagnetic fields in units of the elementary charge ($eE$ and $eB$), which is convenient for our lattice setup. Thus, we will quote results for $g_{a\gamma\gamma} f_a/e^2$. 

\section{Preliminary results}\label{presults}%

\subsection{Lattice setup}\label{latticeset}%

The discretisations used in the simulations were the tree-level improved Symanzik action for the gauge fields and stout-improved staggered quarks for the fermion fields. The electromagnetic fields have been introduced as background fields and we chose the electric field to be imaginary, in order to avoid the sign problem. These background electromagnetic fields are included as extra U(1) phases multiplying the SU(3) links. Since the lattices we are simulating are discretisations of a torus $[L_s^3\times L_t]$, when one considers a closed loop of links there is an ambiguity regarding which is the area that it encloses. By requiring consistency with Stokes' theorem, we obtain a constraint for the possible values of the (imaginary) electric and magnetic fluxes, which need to be integer multiples of $6\pi/(L_sL_t)$ and $6\pi/L_s^2$, respectively (for an explicit derivation, see \cite{fluxquanta}). For the topological charge we have considered two different discretisations, with improvements to quadratic~\cite{qtop} and to quartic order~\cite{qtop_imp} in the lattice spacing. We will refer to these discretisations as `regular' and `improved', respectively. The simulations were carried out at the physical point, using 2 + 1 quark flavours with electric charges $q_u=2e/3$ and $q_d=q_s=-e/3$. 

We have also employed the gradient flow technique to renormalise the lattice gauge fields and to ensure that the topological charge takes on integer values~\cite{luscher2010properties}. For sufficiently high values of the flow time $\tau_f$, we noticed that the observables tend to reach a plateau and that the value at which this occurs increases with the temperature. We have always made sure that a sufficiently high value of $\tau_f$ was used so that plateaus were reached.

\begin{figure}[ht]%
    \centering
    \subfloat{{\includegraphics[width=7.5cm]{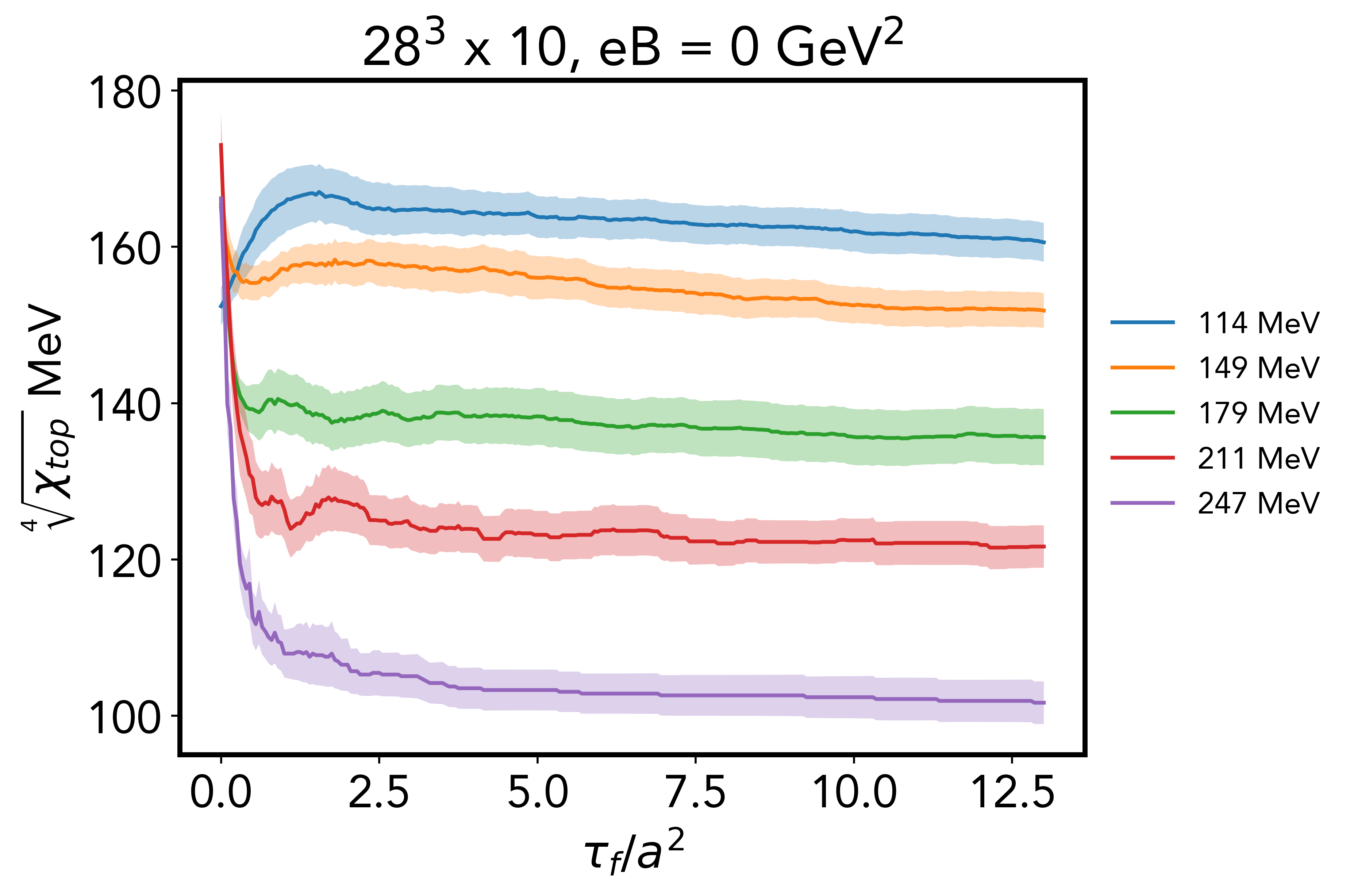} }}%
    \centering
    \subfloat{{\includegraphics[width=6.5cm]{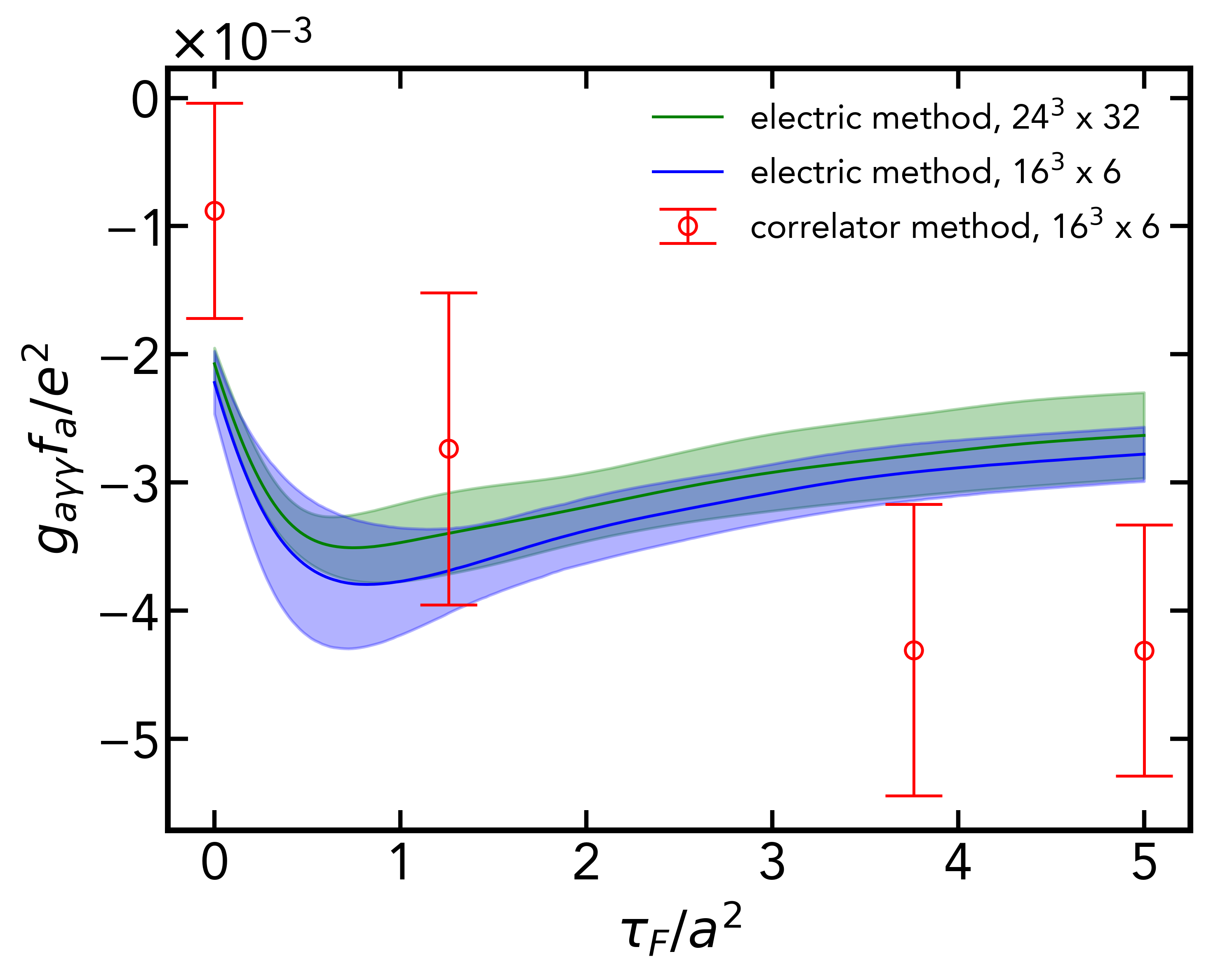} }}%
    \caption{Our observables: the topological susceptibility (left panel) and the axion-photon coupling (right panel) as functions of the Wilson flow parameter. We see that for a large enough amount of flow time, the observables tend to reach a plateau. The improved version of $Q_{\text{top}}$ was used.}%
    \label{wflow}
\end{figure}

\subsection{The topological susceptibility}

\begin{figure}[ht]%
    \centering
    \subfloat{{\includegraphics[width=7cm]{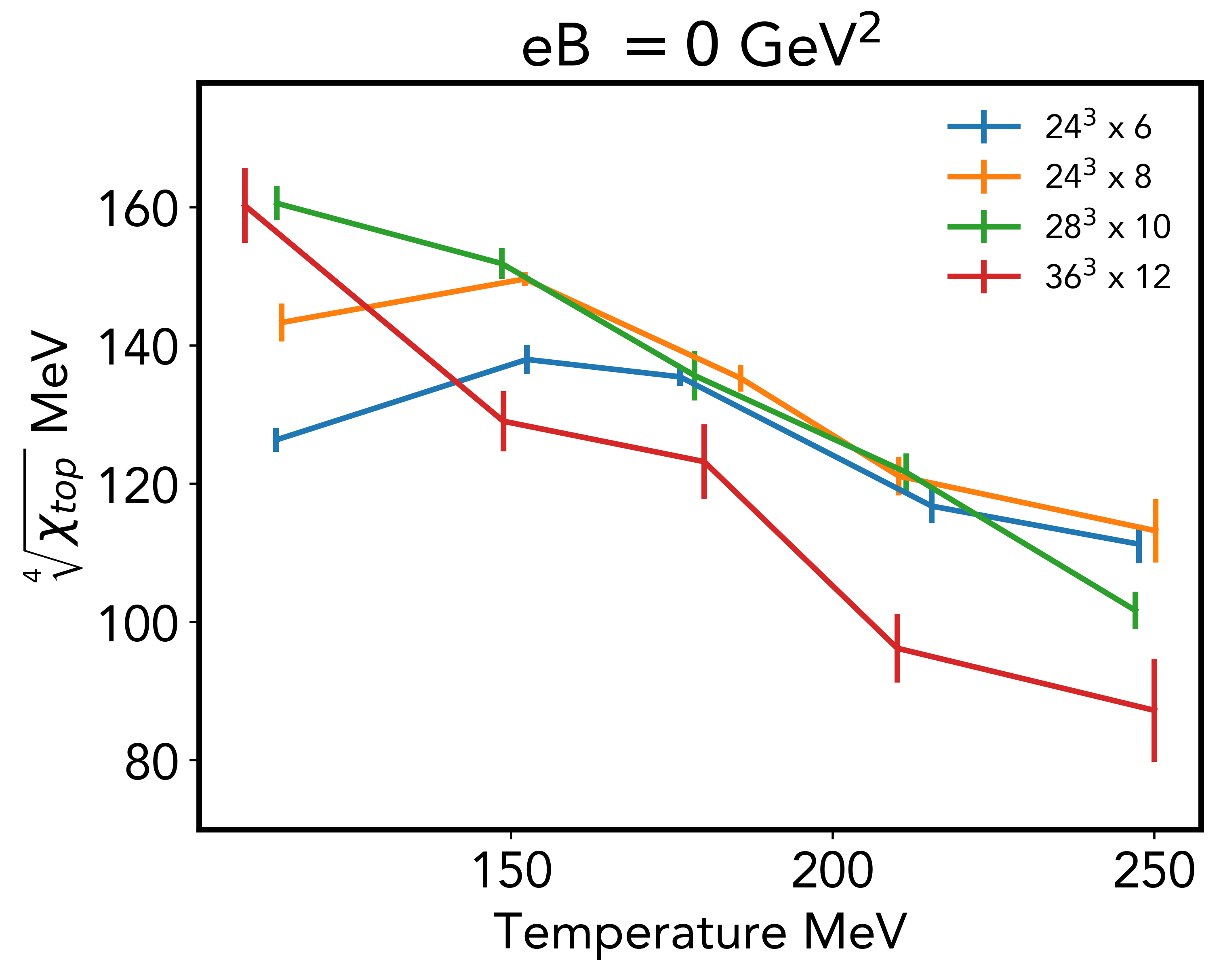} }}%
    \qquad
    \subfloat{{\includegraphics[width=7cm]{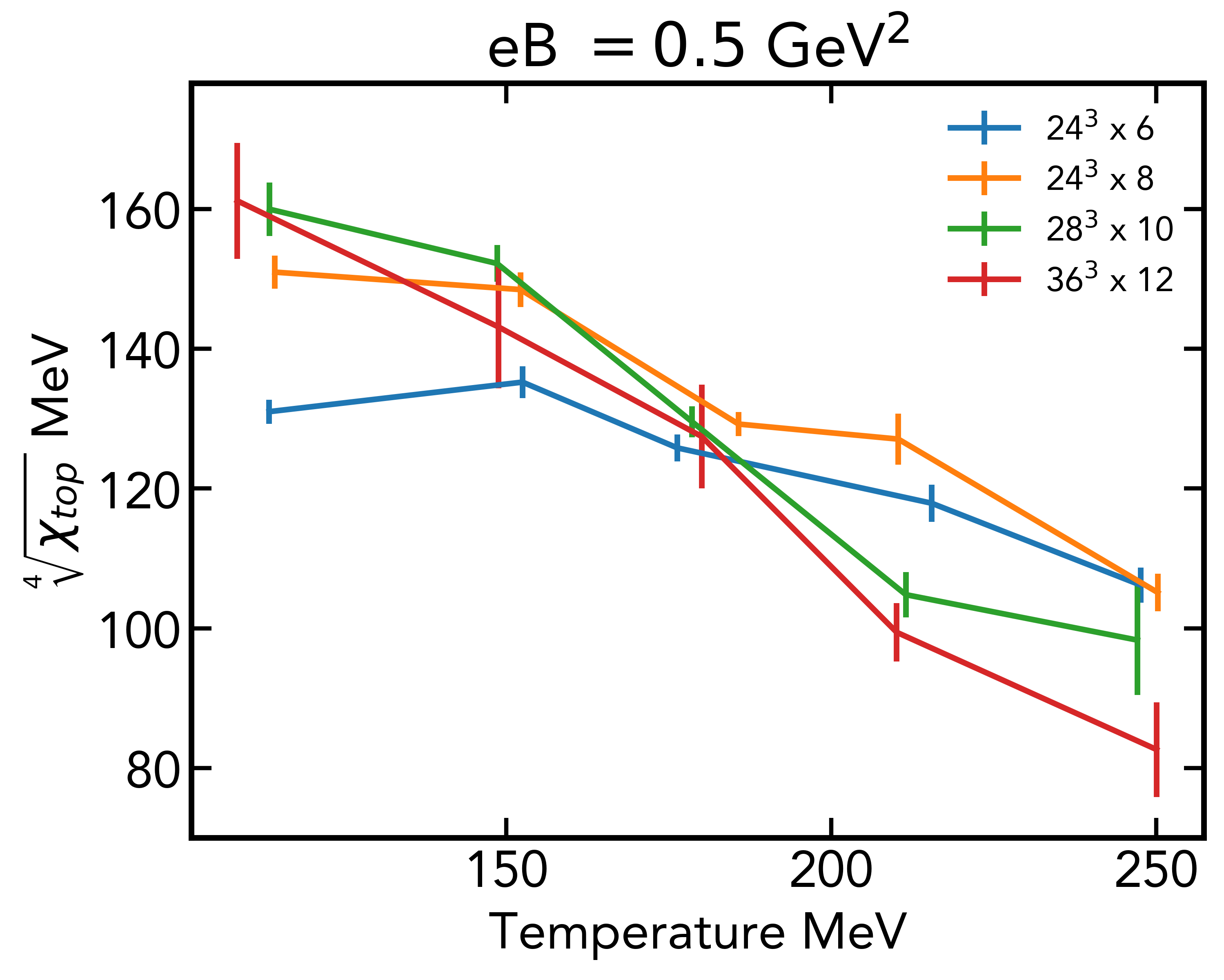} }}\\%
    \centering
    \subfloat{{\includegraphics[width=7cm]{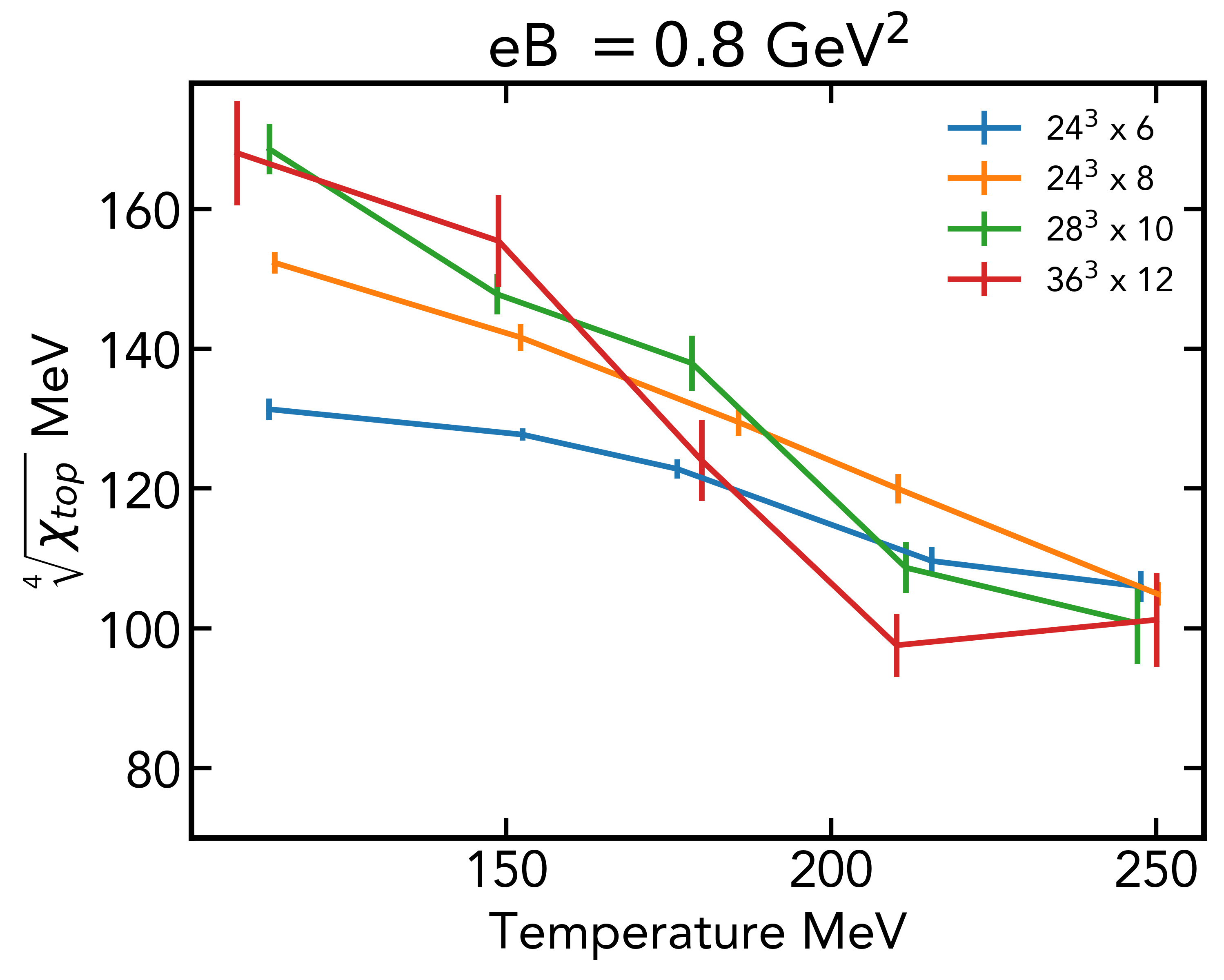} }}%
    \caption{The topological susceptibility as a function of the temperature. Each plot shows a different magnetic field, increasing clockwise. The different colours correspond to different lattice spacings. The improved definition of $Q_{\text{top}}$ was used.}%
    \label{topsusc}
\end{figure}

We have computed the topological susceptibility (axion mass) for a range of temperatures between 110 MeV and 250 MeV. This was done for three different magnetic fields: $eB=0$, 0.5 and 0.8 GeV$^2$. The results are shown in figure \ref{topsusc}, revealing a gradual reduction of $\chi_{\rm top}$ as $T$ grows and mild effects due to $B$. We stress that lattice artefacts are known to enhance $\chi_{\text{top}}$ substantially~\cite{Bonati2016axion,Borsanyi2016lattice}, so that a reliable continuum extrapolation will require finer lattices as well as more involved reweighting approaches. 

\begin{figure}[ht]%
    \centering
    \subfloat{{\includegraphics[width=7cm]{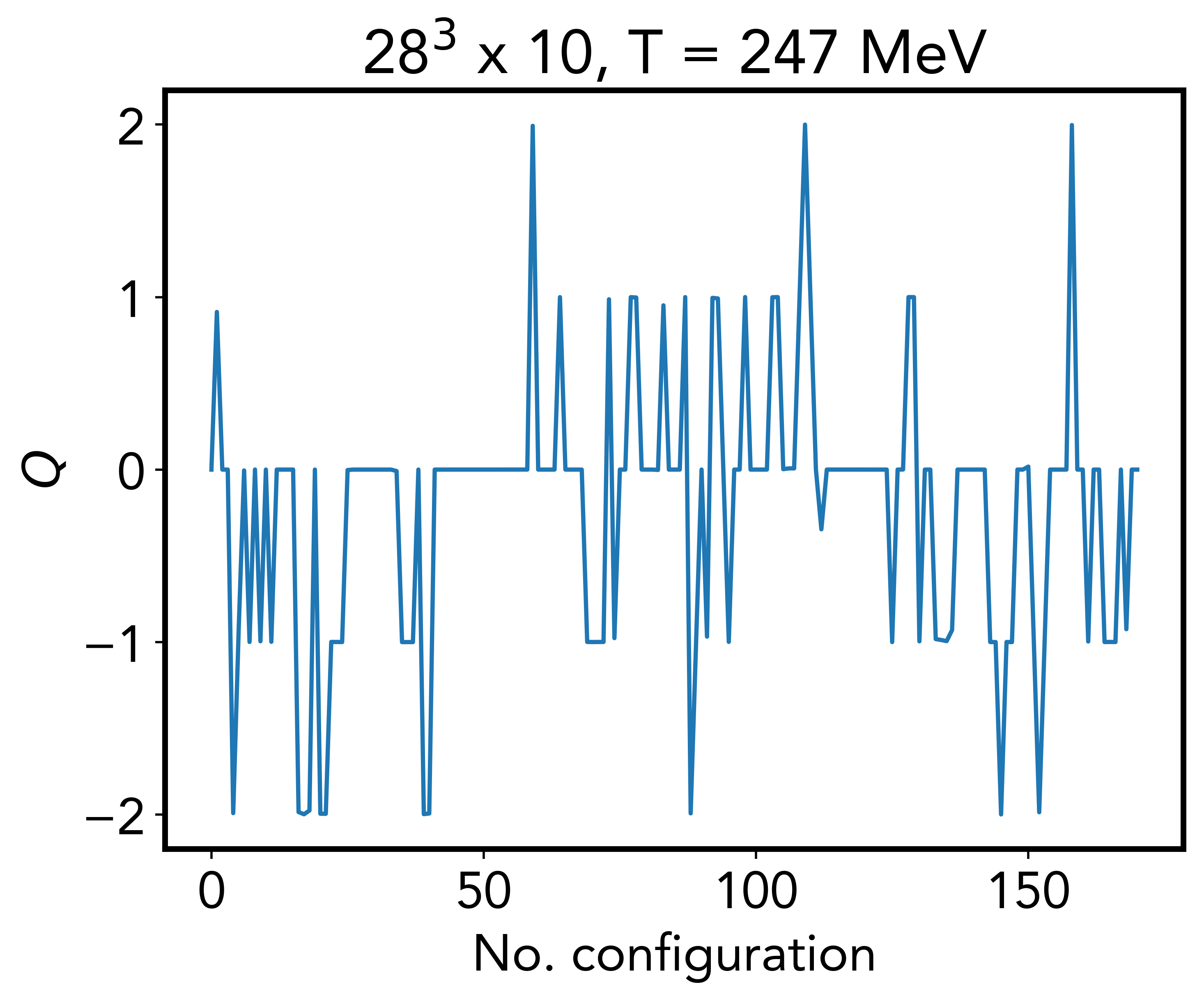} }}%
    \qquad
    \subfloat{{\includegraphics[width=7cm]{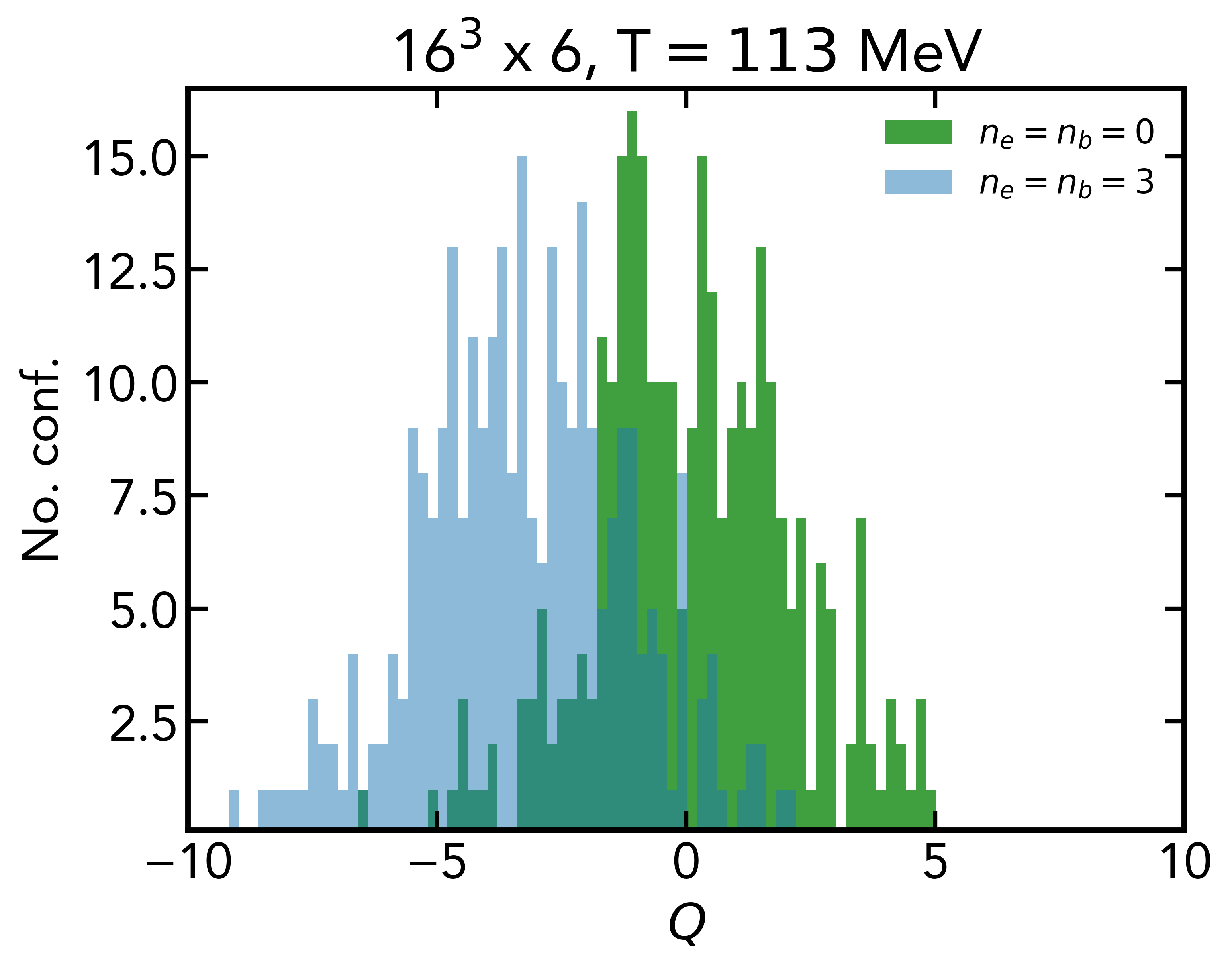} }}%
    \caption{Left: a section of the Monte Carlo history of the topological charge for our $N_t=10$ lattices ($a \approx 0.080$ fm) at zero magnetic field. The improved definition of $Q_{\text{top}}$ was used. Right: comparison between the histogram of the topological charge at zero and nonzero $\Vec{E}\cdot\Vec{B}$. The distribution average shifts from zero to negative values when the electromagnetic fields are  turned on. Here, $n_e$ and $n_b$ denote the electric and magnetic flux quantum numbers. On the left (right) the improved (regular) definition of $Q_{\text{top}}$ was used.}%
    \label{topfreezing}
\end{figure}

Regarding the issue of topological freezing, we have not observed it in the range of lattice spacings used in this study ([0.066-0.29] fm). For a given lattice size, as we increase the temperature we see that the number of topological sectors that are sampled by the Monte Carlo simulation gets reduced and concentrated around zero, but we still notice a sufficient number of fluctuations around the zero sector. We demonstrate this for our $28^3\times10$ ensemble in Fig. \ref{topfreezing}, left panel. 

\subsection{The axion-photon coupling}

\begin{figure}[ht]%
    \centering
    \subfloat{{\includegraphics[width=7cm]{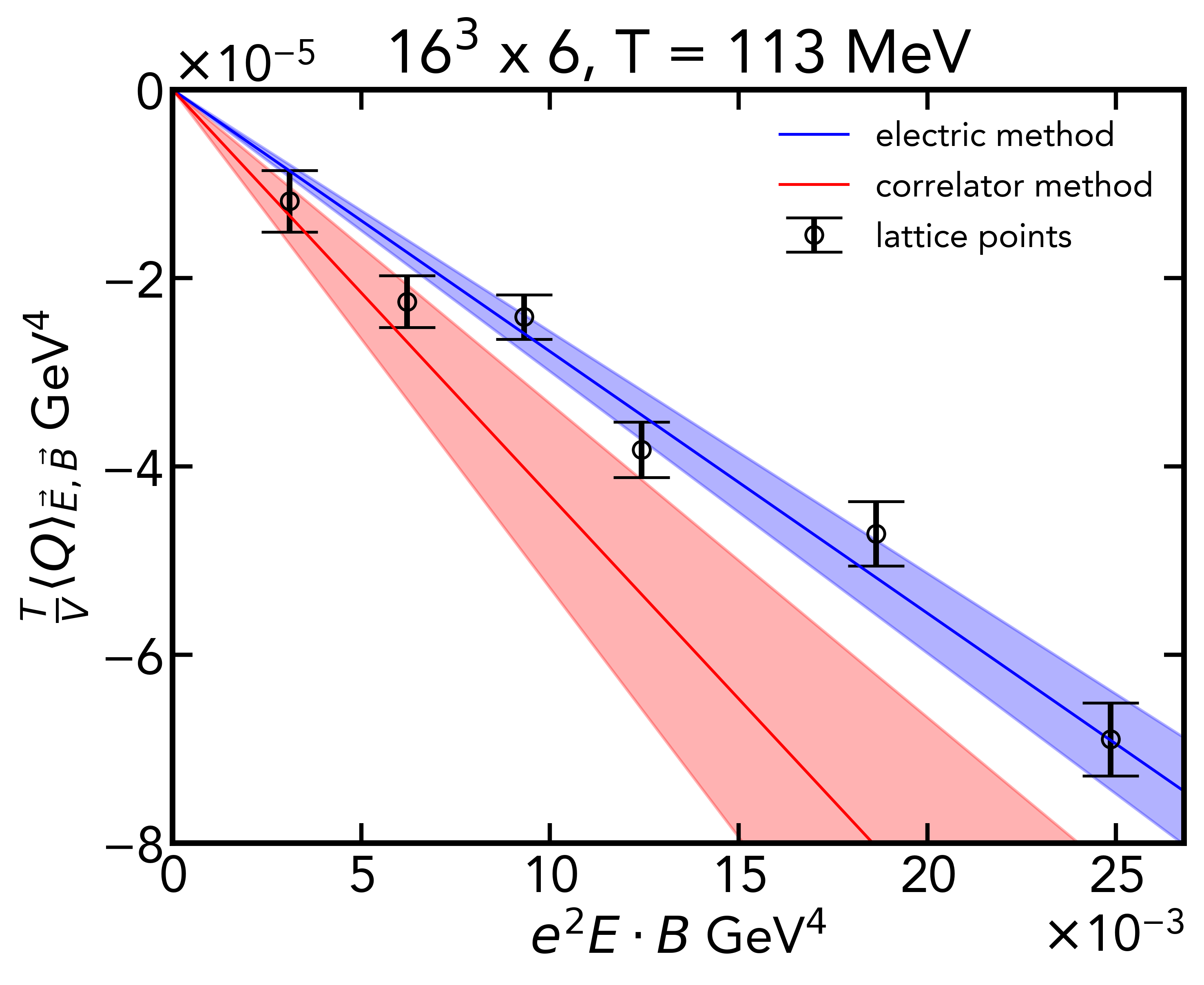} }}%
    \qquad
    \subfloat{{\includegraphics[width=7cm]{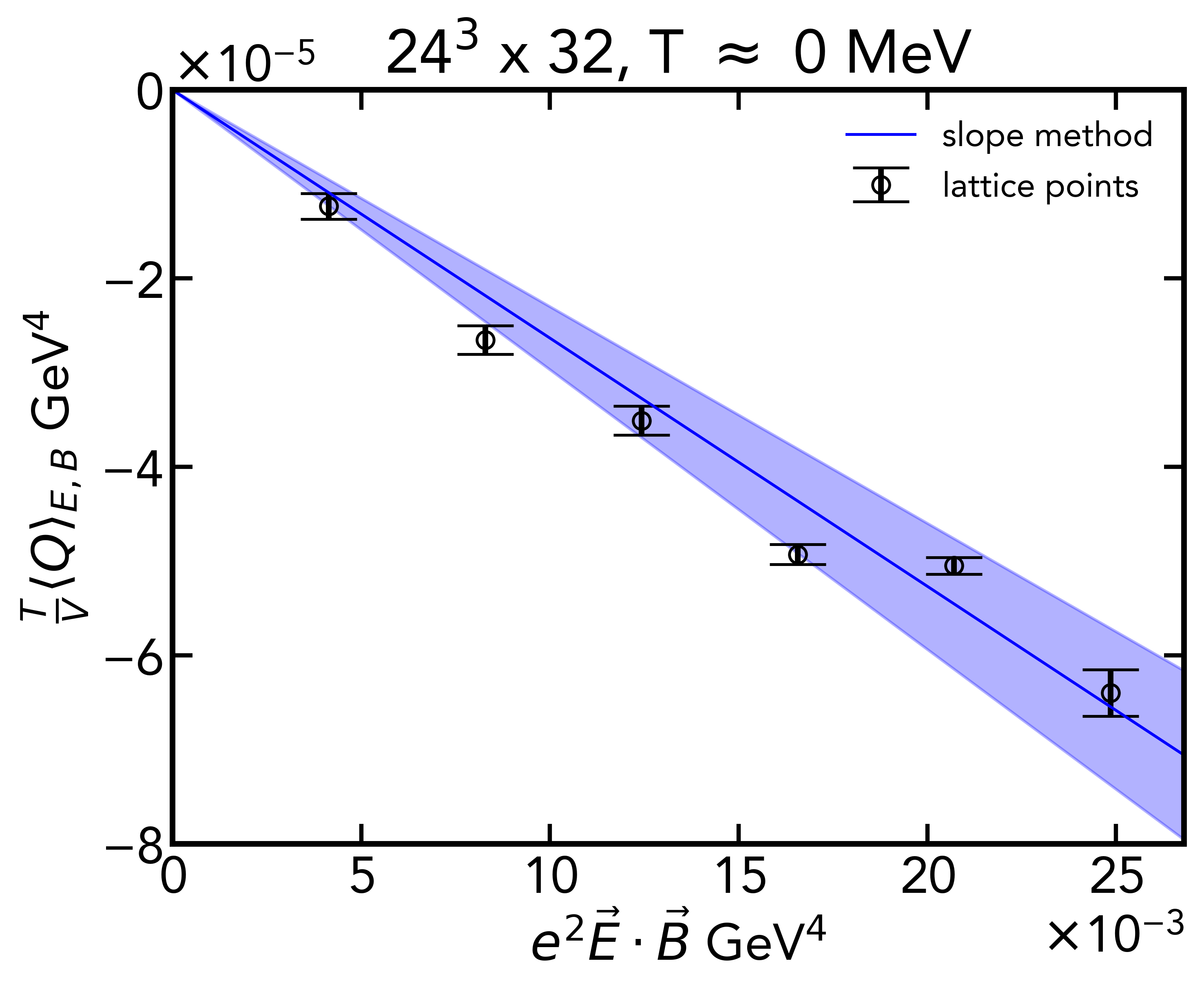} }}%
    \caption{$\langle Q_{\text{top}\rangle}$ as a function of $\Vec{E}\cdot\Vec{B}$ for our $16^3\times6$ (left panel) and $24^3\times 32$ ensembles (right panel). The lattice spacing is $a \approx 0.29$ fm in both lattices. Notice that the errors of the correlator method are approximately 4 times bigger as that of electric method (left plot). The improved version of $Q_{\text{top}}$ was used.}%
    \label{ap_coupling}
\end{figure}

As described in Sec.~\ref{topology}, the introduction of non-orthogonal electric and magnetic fields induces a nonzero topological charge in the gluonic sector. We have checked that this is indeed the case, see Fig.~\ref{topfreezing}, right panel, for an example. As anticipated using the Atiyah-Singer index theorem, the shift is indeed towards negative values. Furthermore, the dependence on $\Vec{E}\cdot\Vec{B}$ is linear for sufficiently weak fields ($\leq 0.025$ GeV$^4$); see Fig.~\ref{ap_coupling}. 

Regarding the calculation of the axion-photon coupling, we have compared the electric and correlator methods in the same volume and at the same temperature, obtaining comparable results, see Fig \ref{ap_coupling}. A priori, the main advantage of the correlator method is that we do not need to include an imaginary electric field. Also, it is computationally advantageous since we can use the configurations at finite magnetic field which are already available from previous studies \cite{Bali:2011qj,Bali:2012zg}. However, the required correlators are quite noisy even when calculated with thousands of random estimators. When comparing these two methods with a similar amount of statistics, we observe that the correlator method leads to errors of about four times larger than the electric method. This, plus the fact that calculating the topological charge is relatively cheap in computational time (also when taking into account the generation of new configurations), prompted us to choose the electric method over the correlator one: see Fig.\ref{wflow}, right panel\footnote{We note this situation is similar to the calculation of the spin polarisation $\langle\bar\psi\sigma_{xy}\psi\rangle$ by means of simulations at nonzero $B$ and by analogous correlators \cite{Bali:2020bcn}.}.

We studied the coupling using two different volumes at two different temperatures but at the same lattice spacing, obtaining a very similar value. Hence, we can say that the volume as well as the temperature dependence below $T_c$ is mild.
The preliminary value of the coupling that we obtain from the $24^3\times32$ lattice  simulations is $g_{a\gamma\gamma}f_a/e^2 = -0.0026(3)$, which is about 9 times smaller than the value obtained from ChPT \cite{di2016qcd}. We expect that simulations at smaller lattice spacings will take this value closer to the ChPT prediction.

 The statistical errors for both observables have been computed using a jackknife procedure. We have also included the error of the topological charge not being exactly an integer on the lattice for $\chi_{\text{top}}$ and for $g_{a\gamma\gamma}$ obtained from the electric method. This was done by rounding the topological charge to the nearest integer and adding the difference in quadrature.

\section{Summary and outlook}%

In this proceeding we have shown preliminary results for the magnetic field dependence of the topological susceptibility at finite temperature. We have also presented two different methods to compute the axion-photon coupling using lattice QCD, which we dubbed electric and correlator methods. We have confirmed the linear response of the QCD topological charge to the non-orthogonal background electromagnetic fields. Finally, we demonstrated how both the topological susceptibility and the axion-photon coupling are under control when using the gradient flow.

For further work we are currently in the process of generating more statistics for both observables. We are also working towards the reweighting of the topological susceptibility to mimic the effect of exact zero modes, as well as generating ensembles with finer lattice spacings for the axion-photon coupling. These aspects will allow for a reliable continuum extrapolations for both observables.

\section*{Acknowledgements}
This work was supported by the Deutsche Forschungsgemeinschaft (DFG, German Research Foundation) – project number 315477589 – TRR 211 and by the Helmholtz Graduate School for Hadron and Ion Research (HGS-HIRe for FAIR). The authors are also grateful to Guy D. Moore for fruitful discussions. The
computations in this work were performed on the GPU cluster at Bielefeld University.

\bibliographystyle{utphys} 
\bibliography{skeleton}

\providecommand{\href}[2]{#2}\begingroup\raggedright\begin{thebibliography}{10}

\bibitem{hook2018tasi}
A.~Hook, ``{TASI Lectures on the Strong CP Problem and Axions},'' {\em PoS}
  {\bfseries TASI2018} (2019) 004,
  \href{http://arxiv.org/abs/1812.02669}{{\ttfamily arXiv:1812.02669
  [hep-ph]}}.

\bibitem{Abel:2020pzs}
C.~Abel {\em et~al.}, ``{Measurement of the Permanent Electric Dipole Moment of
  the Neutron},'' \href{http://dx.doi.org/10.1103/PhysRevLett.124.081803}{{\em
  Phys. Rev. Lett.} {\bfseries 124} no.~8, (2020) 081803},
  \href{http://arxiv.org/abs/2001.11966}{{\ttfamily arXiv:2001.11966
  [hep-ex]}}.

\bibitem{Bhattacharya:2021lol}
T.~Bhattacharya, V.~Cirigliano, R.~Gupta, E.~Mereghetti, and B.~Yoon,
  ``{Contribution of the QCD $\Theta$-term to the nucleon electric dipole
  moment},'' \href{http://dx.doi.org/10.1103/PhysRevD.103.114507}{{\em Phys.
  Rev. D} {\bfseries 103} no.~11, (2021) 114507},
  \href{http://arxiv.org/abs/2101.07230}{{\ttfamily arXiv:2101.07230
  [hep-lat]}}.

\bibitem{Alexandrou:2020mds}
C.~Alexandrou, A.~Athenodorou, K.~Hadjiyiannakou, and A.~Todaro, ``{Neutron
  electric dipole moment using lattice QCD simulations at the physical
  point},'' \href{http://dx.doi.org/10.1103/PhysRevD.103.054501}{{\em Phys.
  Rev. D} {\bfseries 103} no.~5, (2021) 054501},
  \href{http://arxiv.org/abs/2011.01084}{{\ttfamily arXiv:2011.01084
  [hep-lat]}}.

\bibitem{Dragos:2019oxn}
J.~Dragos, T.~Luu, A.~Shindler, J.~de~Vries, and A.~Yousif, ``{Confirming the
  Existence of the strong CP Problem in Lattice QCD with the Gradient Flow},''
  \href{http://dx.doi.org/10.1103/PhysRevC.103.015202}{{\em Phys. Rev. C}
  {\bfseries 103} no.~1, (2021) 015202},
  \href{http://arxiv.org/abs/1902.03254}{{\ttfamily arXiv:1902.03254
  [hep-lat]}}.

\bibitem{PhysRevLett.37.8}
G.~'t~Hooft, ``Symmetry breaking through bell-jackiw anomalies,''
  \href{http://dx.doi.org/10.1103/PhysRevLett.37.8}{{\em Phys. Rev. Lett.}
  {\bfseries 37} (Jul, 1976) 8--11}.

\bibitem{Ellis:1978hq}
J.~R. Ellis and M.~K. Gaillard, ``{Strong and Weak CP Violation},''
  \href{http://dx.doi.org/10.1016/0550-3213(79)90297-9}{{\em Nucl. Phys. B}
  {\bfseries 150} (1979) 141--162}.

\bibitem{PhysRevLett.38.1440}
R.~D. Peccei and H.~R. Quinn, ``$\mathrm{CP}$ conservation in the presence of
  pseudoparticles,'' \href{http://dx.doi.org/10.1103/PhysRevLett.38.1440}{{\em
  Phys. Rev. Lett.} {\bfseries 38} (Jun, 1977) 1440--1443}.

\bibitem{PhysRevD.16.1791}
R.~D. Peccei and H.~R. Quinn, ``Constraints imposed by $\mathrm{CP}$
  conservation in the presence of pseudoparticles,''
  \href{http://dx.doi.org/10.1103/PhysRevD.16.1791}{{\em Phys. Rev. D}
  {\bfseries 16} (Sep, 1977) 1791--1797}.

\bibitem{PhysRevLett.43.103}
J.~E. Kim, ``Weak-interaction singlet and strong $\mathrm{CP}$ invariance,''
  \href{http://dx.doi.org/10.1103/PhysRevLett.43.103}{{\em Phys. Rev. Lett.}
  {\bfseries 43} (Jul, 1979) 103--107}.

\bibitem{Shifman:1979if}
M.~A. Shifman, A.~I. Vainshtein, and V.~I. Zakharov, ``{Can Confinement Ensure
  Natural CP Invariance of Strong Interactions?},''
  \href{http://dx.doi.org/10.1016/0550-3213(80)90209-6}{{\em Nucl. Phys. B}
  {\bfseries 166} (1980) 493--506}.

\bibitem{Dine:1981rt}
M.~Dine, W.~Fischler, and M.~Srednicki, ``{A Simple Solution to the Strong CP
  Problem with a Harmless Axion},''
  \href{http://dx.doi.org/10.1016/0370-2693(81)90590-6}{{\em Phys. Lett. B}
  {\bfseries 104} (1981) 199--202}.

\bibitem{di2016qcd}
G.~Grilli~di Cortona, E.~Hardy, J.~Pardo~Vega, and G.~Villadoro, ``{The QCD
  axion, precisely},'' \href{http://dx.doi.org/10.1007/JHEP01(2016)034}{{\em
  JHEP} {\bfseries 01} (2016) 034},
  \href{http://arxiv.org/abs/1511.02867}{{\ttfamily arXiv:1511.02867
  [hep-ph]}}.

\bibitem{Bonati2016axion}
C.~Bonati, M.~D'Elia, M.~Mariti, G.~Martinelli, M.~Mesiti, F.~Negro,
  F.~Sanfilippo, and G.~Villadoro, ``{Axion phenomenology and
  $\theta$-dependence from $N_f = 2+1$ lattice QCD},''
  \href{http://dx.doi.org/10.1007/JHEP03(2016)155}{{\em JHEP} {\bfseries 03}
  (2016) 155}, \href{http://arxiv.org/abs/1512.06746}{{\ttfamily
  arXiv:1512.06746 [hep-lat]}}.

\bibitem{Borsanyi2016lattice}
S.~Bors\'anyi {\em et~al.}, ``{Calculation of the axion mass based on
  high-temperature lattice quantum chromodynamics},''
  \href{http://dx.doi.org/10.1038/nature20115}{{\em Nature} {\bfseries 539}
  no.~7627, (2016) 69--71}, \href{http://arxiv.org/abs/1606.07494}{{\ttfamily
  arXiv:1606.07494 [hep-lat]}}.

\bibitem{Jahn:2018dke}
P.~T. Jahn, G.~D. Moore, and D.~Robaina, ``{$\chi_{\textrm{top}}(T \gg
  T_{\textrm{c}})$ in pure-glue QCD through reweighting},''
  \href{http://dx.doi.org/10.1103/PhysRevD.98.054512}{{\em Phys. Rev. D}
  {\bfseries 98} no.~5, (2018) 054512},
  \href{http://arxiv.org/abs/1806.01162}{{\ttfamily arXiv:1806.01162
  [hep-lat]}}.

\bibitem{Fukushima:2008xe}
K.~Fukushima, D.~E. Kharzeev, and H.~J. Warringa, ``{The Chiral Magnetic
  Effect},'' \href{http://dx.doi.org/10.1103/PhysRevD.78.074033}{{\em Phys.
  Rev. D} {\bfseries 78} (2008) 074033},
  \href{http://arxiv.org/abs/0808.3382}{{\ttfamily arXiv:0808.3382 [hep-ph]}}.

\bibitem{Adhikari:2021lbl}
P.~Adhikari, ``{Topological susceptibility in a uniform magnetic field},''
  \href{http://dx.doi.org/10.1016/j.physletb.2021.136826}{{\em Phys. Lett. B}
  {\bfseries 825} (2022) 136826},
  \href{http://arxiv.org/abs/2103.05048}{{\ttfamily arXiv:2103.05048
  [hep-ph]}}.

\bibitem{DElia:2012ifm}
M.~D'Elia, M.~Mariti, and F.~Negro, ``{Susceptibility of the QCD vacuum to
  CP-odd electromagnetic background fields},''
  \href{http://dx.doi.org/10.1103/PhysRevLett.110.082002}{{\em Phys. Rev.
  Lett.} {\bfseries 110} no.~8, (2013) 082002},
  \href{http://arxiv.org/abs/1209.0722}{{\ttfamily arXiv:1209.0722 [hep-lat]}}.

\bibitem{indexth}
D.~S. Freed, ``{The Atiyah\textendash{}Singer index theorem},''
  \href{http://dx.doi.org/10.1090/bull/1747}{{\em Bull. Am. Math. Soc.}
  {\bfseries 58} no.~4, (2021) 517--566},
  \href{http://arxiv.org/abs/2107.03557}{{\ttfamily arXiv:2107.03557
  [math.HO]}}.

\bibitem{Endrodi:2021qxz}
G.~Endr\H{o}di and G.~Mark\'o, ``{Thermal QCD with external imaginary electric
  fields on the lattice},'' \href{http://dx.doi.org/10.22323/1.396.0245}{{\em
  PoS} {\bfseries LATTICE2021} (2022) 245},
  \href{http://arxiv.org/abs/2110.12189}{{\ttfamily arXiv:2110.12189
  [hep-lat]}}.

\bibitem{Endrodi:2022wym}
G.~Endr\H{o}di and G.~Mark\'o, ``{On electric fields in hot QCD: perturbation
  theory},'' \href{http://arxiv.org/abs/2208.14306}{{\ttfamily arXiv:2208.14306
  [hep-ph]}}.

\bibitem{Bali:2014vja}
G.~S. Bali, F.~Bruckmann, G.~Endr\H{o}di, Z.~Fodor, S.~D. Katz, and
  A.~Sch{\"a}fer, ``{Local CP-violation and electric charge separation by
  magnetic fields from lattice QCD},''
  \href{http://dx.doi.org/10.1007/JHEP04(2014)129}{{\em JHEP} {\bfseries 04}
  (2014) 129}, \href{http://arxiv.org/abs/1401.4141}{{\ttfamily arXiv:1401.4141
  [hep-lat]}}.

\bibitem{Bali:2020bcn}
G.~S. Bali, G.~Endr\H{o}di, and S.~Piemonte, ``{Magnetic susceptibility of QCD
  matter and its decomposition from the lattice},''
  \href{http://dx.doi.org/10.1007/JHEP07(2020)183}{{\em JHEP} {\bfseries 07}
  (2020) 183}, \href{http://arxiv.org/abs/2004.08778}{{\ttfamily
  arXiv:2004.08778 [hep-lat]}}.

\bibitem{fluxquanta}
M.~H. Al-Hashimi and U.~J. Wiese, ``{Discrete Accidental Symmetry for a
  Particle in a Constant Magnetic Field on a Torus},''
  \href{http://dx.doi.org/10.1016/j.aop.2008.07.006}{{\em Annals Phys.}
  {\bfseries 324} (2009) 343--360},
  \href{http://arxiv.org/abs/0807.0630}{{\ttfamily arXiv:0807.0630
  [quant-ph]}}.

\bibitem{qtop}
S.~Bors\'anyi {\em et~al.}, ``{High-precision scale setting in lattice QCD},''
  \href{http://dx.doi.org/10.1007/JHEP09(2012)010}{{\em JHEP} {\bfseries 09}
  (2012) 010}, \href{http://arxiv.org/abs/1203.4469}{{\ttfamily arXiv:1203.4469
  [hep-lat]}}.

\bibitem{qtop_imp}
S.~O. Bilson-Thompson, D.~B. Leinweber, and A.~G. Williams, ``{Highly improved
  lattice field strength tensor},''
  \href{http://dx.doi.org/10.1016/S0003-4916(03)00009-5}{{\em Annals Phys.}
  {\bfseries 304} (2003) 1--21},
  \href{http://arxiv.org/abs/hep-lat/0203008}{{\ttfamily
  arXiv:hep-lat/0203008}}.

\bibitem{luscher2010properties}
M.~L\"uscher, ``{Properties and uses of the Wilson flow in lattice QCD},''
  \href{http://dx.doi.org/10.1007/JHEP08(2010)071}{{\em JHEP} {\bfseries 08}
  (2010) 071}, \href{http://arxiv.org/abs/1006.4518}{{\ttfamily arXiv:1006.4518
  [hep-lat]}}. [Erratum: JHEP 03, 092 (2014)].

\bibitem{Bali:2011qj}
G.~S. Bali, F.~Bruckmann, G.~Endr\H{o}di, Z.~Fodor, S.~D. Katz, S.~Krieg,
  A.~Sch{\"a}fer, and K.~K. Szab\'o, ``{The QCD phase diagram for external
  magnetic fields},'' \href{http://dx.doi.org/10.1007/JHEP02(2012)044}{{\em
  JHEP} {\bfseries 02} (2012) 044},
  \href{http://arxiv.org/abs/1111.4956}{{\ttfamily arXiv:1111.4956 [hep-lat]}}.

\bibitem{Bali:2012zg}
G.~S. Bali, F.~Bruckmann, G.~Endr\H{o}di, Z.~Fodor, S.~D. Katz, and
  A.~Sch{\"a}fer, ``{QCD quark condensate in external magnetic fields},''
  \href{http://dx.doi.org/10.1103/PhysRevD.86.071502}{{\em Phys. Rev. D}
  {\bfseries 86} (2012) 071502},
  \href{http://arxiv.org/abs/1206.4205}{{\ttfamily arXiv:1206.4205 [hep-lat]}}.

\end{thebibliography}\endgroup

\end{document}